\pdfoutput=1
\documentclass[aps,prb,amsmath,amssymb,twocolumn,letterpaper,%
         footinbib,bibnotes,reprint,nobalancelastpage,raggedbottom,showpacs,
         citeautoscript,floatfix]{revtex4-1}
\usepackage[usenames,dvipsnames,table]{xcolor}
\usepackage[
  breaklinks,             %
  bookmarks = false, %
  pdfpagemode   = UseNone,%
  pdfstartview 	= FitH,		%
  pdfstartpage  = 1,      %
  colorlinks    = true,   %
]{hyperref}
\hypersetup{
    linkcolor=Magenta,          
    citecolor=Emerald,        
    filecolor=Mulberry,      
    urlcolor=CornflowerBlue           
}
\usepackage{microtype}
\usepackage{graphicx}
\usepackage[version=3]{mhchem}
\usepackage{textcomp}
\usepackage{amssymb}



\begin{document}
\title{Octahedral Engineering of Orbital Polarizations in Charge Transfer Oxides}
\author{Antonio Cammarata}
	\email{acammarata@coe.drexel.edu}
	\affiliation{Department\! of\! Materials\! Science\! \& Engineering,\!
	Drexel University,\! Philadelphia,\! PA 19104,\! USA}%
\author{James M.\ Rondinelli}
  \email{jrondinelli@coe.drexel.edu}
	\affiliation{Department\! of\! Materials\! Science\! \& Engineering,\!
	Drexel University,\! Philadelphia,\! PA 19104,\! USA}%
\date{\today}
\begin{abstract}
Negative charge transfer $AB$O$_3$ oxides may undergo 
electronic metal--insulator transitions (MIT) concomitant with a dilation and 
contraction of nearly rigid octahedra.
On both sides of the MIT are in-phase or out-of-phase (or both) rotations of 
adjacent octahedra that buckle the 
$B$--O--$B$ bond angle away from 180$^\circ$.
%
Using density functional theory with the PBEsol$+U$ approach, 
we describe a novel octahedral engineering avenue to control the $B$ $3d$ and O $2p$ 
orbital polarization through enhancement of the $B$O$_6$ rotation ``sense'' rather than solely through conventional 
changes to the $B$--O bond lengths, \emph{i.e.} crystal field distortions.
%
%
Using CaFeO$_3$ as a prototypical material, we show the flavor of the octahedral rotation pattern when combined with strain--rotation coupling and thin film engineering strategies offers a promising avenue to fine tune orbital polarizations near electronic phase boundaries.
\end{abstract}
\maketitle
\sloppy
\section{Introducion}

Transition metal oxide (TMO) perovskites are known to be strongly correlated materials \cite{Dagotto:2005}, whose properties are controlled by a complex interplay between geometric and electronic degrees of freedom.
These are determined by considering the relative magnitude between various energy scales and interactions: 
the energy difference of the transition metal ($M$) $d$ orbitals and the 
oxygen $p$ states, referred to as the charge transfer energy, 
and the strength of the on-site Hubbard $U$ interaction of the $d$ electrons.
The charge transfer energy is more important in TMO where low-energy 
excitations are of $p\!-\!d$-type, whereas the Coulombic interaction, which localizes the electrons on the $M$-site, produces the insulating state in Mott-Hubbard 
systems \cite{Imada/Fujimori/Tokura:1998}.
The properties of correlated electrons of TMO are controlled in part by the relative occupancy of the different transition-metal $d$ orbitals\cite{TNs00_462}. The relative $d$ orbital occupancy is largely determined by the crystal field experienced by the transition-metal cation; this electrostatic field is the result of the 2$p$ electronic density of the coordinating oxygen ligands. The latter, in turn, is directed by the \emph{extended} geometric arrangement of the nearest neighboring oxygen atoms.

Orbital occupancy can be tuned, for example, through chemical substitution \cite{Salamon/Marcelo:2001,Werner/Millis:2007}, epitaxial strain \cite{0295-5075-96-5-57004,Moon/Chakhalian/Rondinelli:2012}, or by superlattice formation in thin films \cite{PhysRevB.82.134408,PhysRevB.83.161102,PhysRevB.82.235123}.
Isovalent substitutions are important to charge transfer-type 
oxides, because rather than modifying the $M$-site 
electronic configuration, cations with different ionic sizes but the same formal 
valence renormalize the transfer interaction and 
the one-electron bandwidths  through changes in the crystal structure, 
\emph{i.e.}, interatomic bond angles and distances.
The crucial distortion in $AB$O$_3$ perovskites is the buckling of the $B$--O--$B$ bond angles, because the 
effective $d$ electron transfer interaction between the neighboring transition metal sites 
is mediated by the \emph{angular overlap} with the O 2$p$ states \cite{doi:10.1021/ic50146a042}.
When the $B$--O--$B$ bond angle deviates from the ideal value of 
180$^\circ$, the transfer interaction weakens and the  
bandwidth narrows.
Such distortions to the inter-octahedral bond angles are typical in GdFeO$_3$-type
perovskites, and therefore knowing how to control the \emph{amplitude} 
of the bond angles distortions is critical to tailoring the charge transfer and $d$-orbital polarization.
While it is well-established that greater octahedral rotations produce more buckled 
bond angles and narrower bandwidth
\cite{Lprb11_235136}, 
it is not well-understood how the rotation ``sense'' --- be it 
in-phase or out-of-phase along a specific 
Cartesian direction ---  influences the differential charge 
occupancy on the $B$-site $d$ and O $2p$ orbitals.

In this work, we use density functional calculations to show 
that controlling the subtle flavor of the octahedral rotation sense 
is as important as the overall amplitude of the rotations 
when engineering the electronic structure, \emph{vis-\`a-vis} orbital polarization, of charge 
transfer oxides near electronic MIT.
We explain this behavior using the definition of \emph{orbital polarization} 
$\mathcal{P}$ of $m_{l1}$ orbital relative to $m_{l2}$ orbital, 
\begin{equation}
\label{eq:pol}
 \mathcal{P}_{l_1 m_{l1},l_2 m_{l2}}=\frac{n_{l_1 m_{l1}}-n_{l_2 m_{l2}}}{n_{l_1 m_{l1}}+n_{l_2 m_{l2}}}\, , 
\end{equation}
where $n_{l_1 m_{l1}}$ and $n_{l_2 m_{l2}}$ are the occupancies of $\left|l_1 m_{l1}\right\rangle$ and $\left|l_2 m_{l2}\right\rangle$ orbitals, with orbital quantum number $l_i$ 
and magnetic quantum number $m_{li}$,  
respectively \cite{PhysRevB.82.134408}.  
With this definition, orbital polarization becomes an effective measure of the 
charge  excess in the former orbital with respect to the latter. 
The metric then enables us to isolate the contribution of each rotation sense to 
the orbital polarization by judicious choice of structural distortions and subsequent 
calculation of $\mathcal{P}$. 
We demonstrate this utility for the specific case of the 
Fe $3d$ ($e_g$-symmetry) and O $2p$ orbitals in CaFeO$_3$.
Specifically, we show 
that out-of-phase octahedral rotations drive a transfer of electronic 
charge to the apical $2p$ orbitals along the Fe--O chains collinear with the axes 
of rotation, whereas the Fe $e_g$ state are less sensitive. 
We then use tensile epitaxial strain as a practical handle to enhance the orbital polarization
using the $\mathcal{P}$-contribution from the out-of-phase octahedral rotations 
to tune the ligand charge density---a crucial parameter 
involved in oxide-based MIT materials.
Our results indicate that electronic-structure engineering requires both the 
amplitude and sense of the octahedral rotations be considered on equal 
footing when designing perovskite TMO for integration into 
next-generation electronics \cite{ha:071101}.
\section{Structural and Computational details}

To explore the rotation--orbital polarization interactions, we make use of the prototypical charge 
transfer oxide \ce{CaFeO3} (CFO), because
%
it lacks Jahn-Teller distortions \cite{TOOAKTjac01_27,WKVPic02_1920} 
and exhibits a sequence of structural phase transitions 
concomitant with a first-order MIT near room temperature (RT) 
\cite{PITLMprb99_10788,Mjpcm05_753,*Mprl05_137205}.
Upon cooling below 290 K, the metallic paramagnet
becomes semiconducting (space group $P2_1/n$) \cite{SPSprb05_45143}, whereby the appearance of an octahedral breathing distortion (BD) together with the octahedral rotations, 
acts to open the electronic gap\cite{CRprb12_195144}, as it has also been observed in rare-earth nickelates \cite{PhysRevB.84.165119,*PhysRevLett.82.3871,*PhysRevLett.109.156402,*PhysRevB.85.214431}.

\begin{figure}[b]
  \centering
  \includegraphics[width=0.27\textwidth,clip]{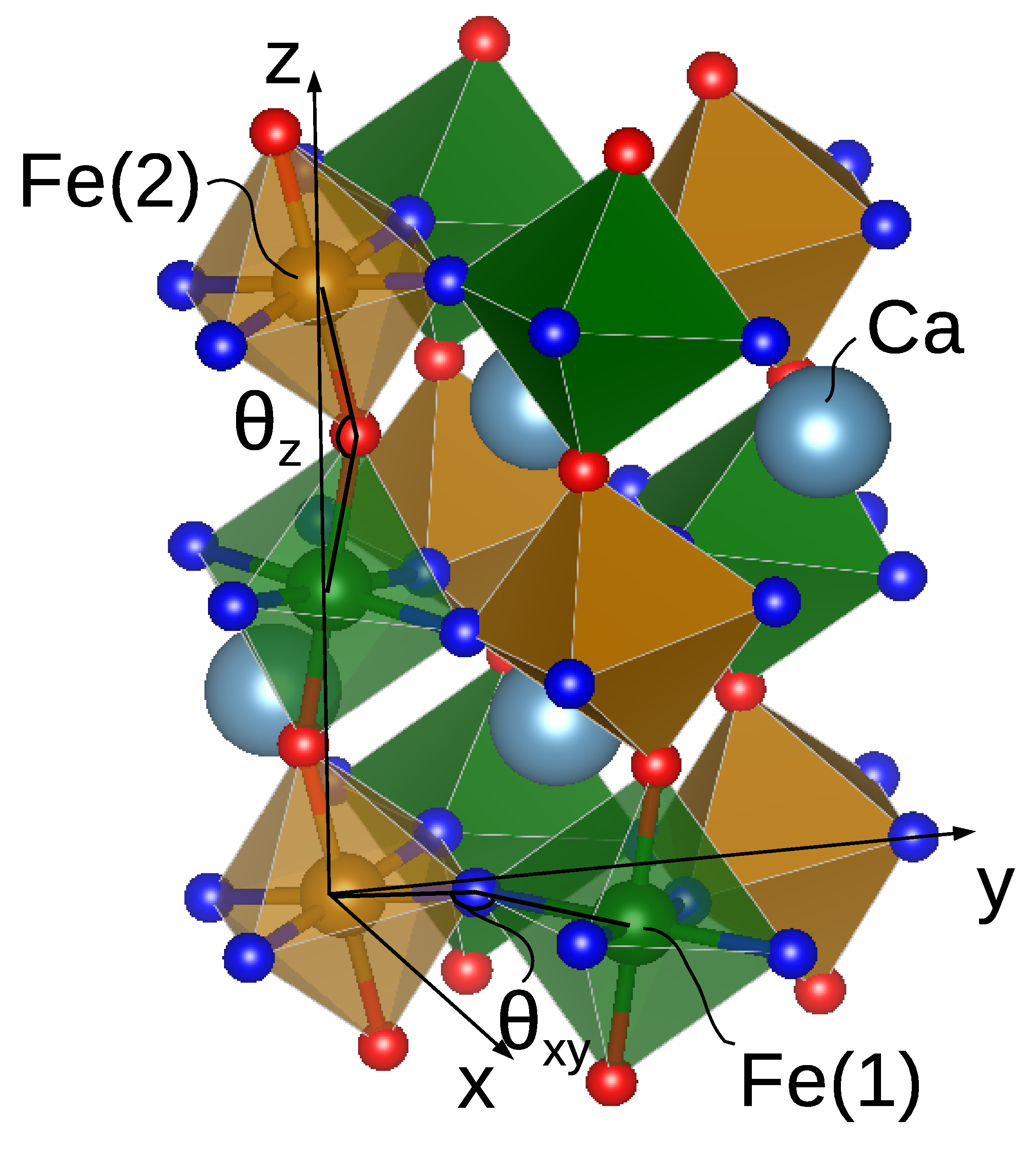}\vspace{-6pt}
  \caption{\label{fig:rot_ang} Rotationally distorted \ce{CaFeO3} structure with the $8d$ (equatorial) and $4c$ (apical) oxygen Wyckoff positions indicated by blue and red spheres, respectively. Pseudo-cubic axes ($x$, $y$, and $z$) and Fe--O--Fe bond  angles $\theta_{xy}$ and $\theta_z$ are also shown.}
\end{figure}

Besides the octahedral breathing distortion, which consists of uniform 
elongation and contraction of Fe--O bond lengths of adjacent octahedra, 
CaFeO$_3$ adopts \footnote{Strictly, 
the low-temperature insulating monoclinic phase adopts the $a^-b^-c^+$, and 
thus the angle of rotations about each axis are unique, but these deviations in the bond angles 
are small.} the most common $a^-a^-c^+$ Glazer tilt pattern \cite{Glazer:1972}, with buckled 
Fe--O--Fe bond angles 
(\autoref{fig:rot_ang}).
Glazer notation  allows us to distinguish the relative sense of the 
octahedral rotations about  Cartesian axes passing through the Fe cation.
The $+$ ($-$) superscript indicates in-phase (out-of-phase) rotations about each 
$x$-, $y$- and $z$-axis. This rotation pattern consists of in-phase rotations about the 
$z$-axis ($a^0a^0c^+$) and is accompanied by two out-of-phase rotations, $a^-a^0c^0$ and 
$a^0a^-c^0$, of the same magnitude about the $x$- and $y$-axes, respectively.
%
%
%

For a perovskite with a Bravais lattice that has orthogonal interaxial angles and the 
$a^-a^-c^+$ pattern, the two important Fe--O--Fe bond angles are $\theta_{xy}$ and $\theta_z$ (\autoref{fig:rot_ang}).
When only in-phase $a^0a^0c^+$ or out-of-phase $a^0a^0c^-$ rotation patterns are present, the $\theta_{xy}$ Fe--O--Fe angle measured within two subsequent Fe--O layers orthogonal to the $z$-direction is the same.
The phase, \emph{i.e.}, the ``rotation sense,'' does not change the rotation magnitude across the layers.
%
Na\"ively, one would deduce that the electronic structure, and in turn the orbital polarizations, is insensitive to the sense of the rotations, since both rotations induce  the same rotation magnitude about the $z$-axis---we investigate this supposition explicitly.
To evaluate how the rotation sense alters the electronic distribution, we carry out density functional 
calculations using the  projector-augmented wave (PAW)
formalism \cite{Bprb94_17953}  
as implemented in the Vienna \emph{Ab initio} Simulation Package (VASP) 
\cite{KFcms96_15,*KJprb99_1758}  within the revised Perdew-Burke-Ernzerhof (PBE) 
generalized gradient approximation  for densely packed solids  
\cite{PBE96_3865,*PRCVSCZBprl_08136406} plus Hubbard $U$ method 
\cite{AALjpcm97_767} (PBEsol$+U$) based on the electronic and atomic 
structural analysis reported in our previous study \cite{CRprb12_195144}. 
We chose the spherically averaged form of the 
rotationally invariant effective $U$ parameter  of 
Dudarev \emph{et al.}\ \cite{DBSHSprb98_1505} with a $U_\textrm{eff}=3.0$~eV, 
hereafter $U$, on the Fe $d$ orbitals.
Note that no significant differences in the main features of the MIT 
occur for values up to $U$ values of 4.0 eV, as previously reported in literature \cite{SPSprb05_45143}, and we  treat the double-counting term within the fully localized limit 
\cite{PhysRevB.79.035103}.
We impose FM order on all Fe-sites then fully relax the spin density.
We use a plane wave energy cutoff of 600 eV, which is then increased up to 850 eV for calculations of the electronic densities of states (DOS).
Ground state structures (DOS) are obtained by sampling the 
Brillouin zone with a minimum of a 
$7\times7\times7$ ($9\times9\times7$) $k$-point mesh and the 
integrations performed with 20~meV Gaussian smearing (tetrahedron method).
Atomic position relaxation are stopped when Hellmann-Feynman forces are minimized to a 0.5 meV\! \AA$^{-1}$ 
tolerance.

%

\section{Results and discussion}
\begin{figure}[t]
  \centering
  \includegraphics[width=0.46\textwidth,clip]{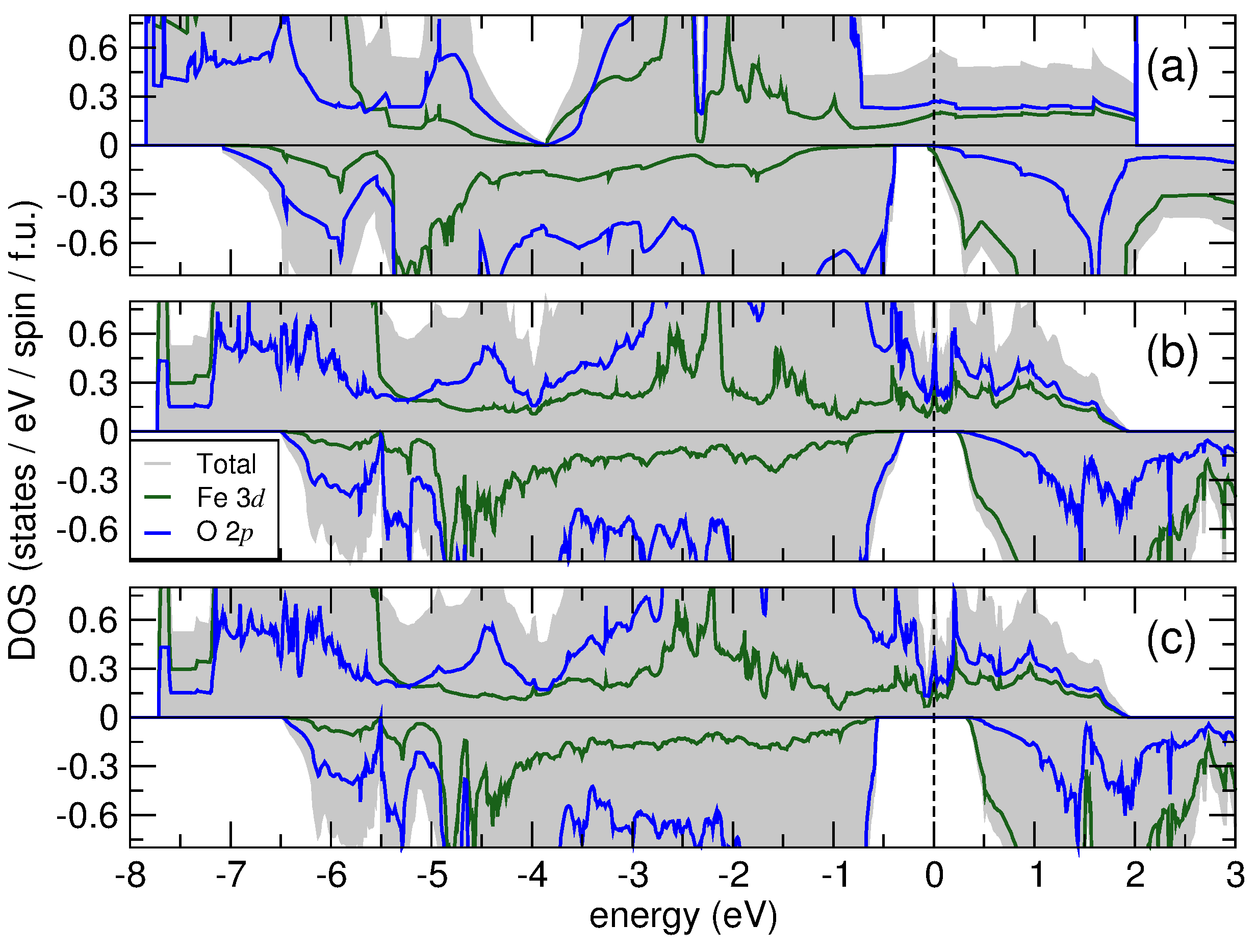}\vspace{-10pt}
  \caption{\label{fig:dos} Total and atom-resolved CaFeO$_3$ density of states (DOS) for structures with (a) $Pm\bar{3}m$, (b) $P4/nmc$ and (c) $I4/m$ symmetry. The structures in (b) and (c) are obtained by applying a  breathing distortion together with $a^0a^0c^+$ and $a^0a^0c^-$ rotations  to the cubic structure, respectively, at identical  breathing amplitudes and $\theta_{xy}$ rotation angles but different rotation sense (cf.\ filled circles in \autoref{fig:pol}). Differences at the Fermi level (0.0~eV) induced by the two rotation senses are nearly indistinguishable.
}
\end{figure}

Our starting point is the analysis of the electronic DOS of cubic \ce{CaFeO3} without any octahedral rotations [\autoref{fig:dos}(a)]. The features of interest are in the energy range -8 to 3~eV. The O 2$p$ states dominate the entire range, validating the charge transfer description. An energy gap appears at the Fermi level ($E_F$) in the spin-down channel. The Fe 3$d$ orbitals contribute to the spin-down states with a main band centered near -5 eV. In the spin-up channel, the Fe 3$d$ states are divided into two principal bands centered about -6.6 and -2.2 eV, respectively, contributing to the metallicity with a broad band extending up to 2 eV above $E_F$.

%

We now examine the symmetry reduction consequences from the octahedral rotations 
and breathing distortions (BD) on the electronic structure of the cubic phase. The BD mode is responsible for splitting of the $1a$ Fe site in the undistorted structure ($Pm\bar{3}m$) into two Wyckoff positions, $2b$ and $2c$, each centered in smaller and larger octahedra arranged as in a 3D checkerboard (\autoref{fig:rot_ang}).
Both the in-phase $a^0a^0c^+$ and the out-of phase $a^0a^0c^-$ 
rotations split the $3d$ oxygen site symmetries in the parent phase into two distinct sets; the consequence of these distortions is that the Fe $d$-orbitals experience an electrostatic interaction from the oxygen ligands which no longer has  
the complete octahedral cubic symmetry. Coupling the BD with in-phase or out-of-phase rotations, results in structures with $P4/nmc$ or $I4/m$ tetragonal symmetry, respectively.
This symmetry lowering changes the DOS with respect to the cubic phase mainly in energy windows far from $E_F$: The spin-down bandwidth decreases, enhancing the band gap,   
and the spin-up states shift 
to higher energies 
in the conduction band (near 1 eV). 
Comparison of the DOS for the different rotation flavors 
[\autoref{fig:dos}(b) and (c)], however, shows no appreciable differences
near $E_F$; the low energy bands appear insensitive to the rotation sense.


\subsection{Orbital Polarization --- $a^0a^0c^{+/-}$ rotations}
\begin{figure}
\flushleft
\hspace*{-4pt}  \includegraphics[width=1.01\columnwidth,clip]{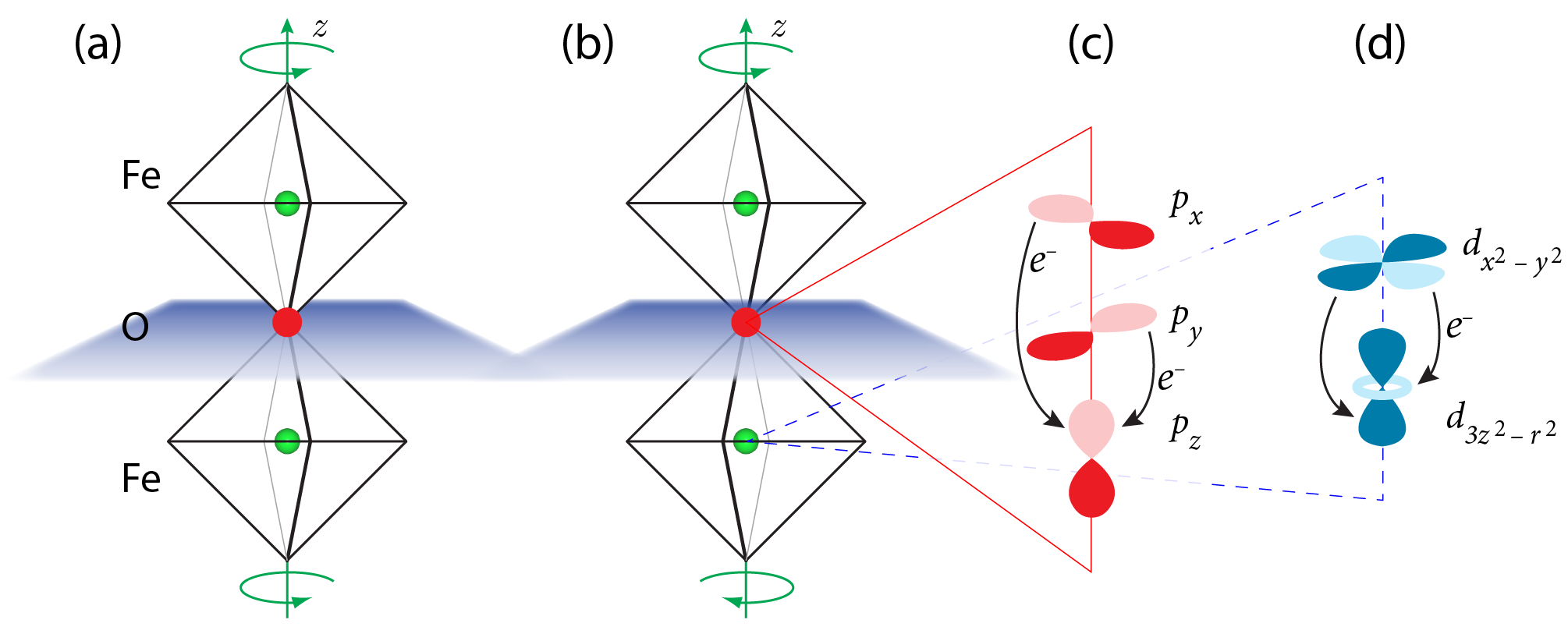}\vspace{-6pt}
  \caption{\label{fig:scheme} (a) $a^0a^0c^+$ and (b) $a^0a^0c^-$ rotation patterns of two adjacent octahedra along the $z$-axis. (c) The out-of-phase rotations induce a transfer of charge from the equatorial plane (shaded), containing the $p_x$ and $p_y$ orbitals of the apical oxygen atom, to the Fe--O apical bond about the $z$-axis, increasing the electron occupancy of the $p_z$ orbital lying along it. (d) On the Fe site, irrespective of the rotation sense, octahedral rotations induce charge transfer from the $d_{x^2-y^2}$ to the $d_{3z^2-r^2}$ orbital, the latter directed along the apical Fe--O bond.
}
\end{figure}

\begin{figure*}[t]
  \centering
  \includegraphics[width=0.75\textwidth,clip]{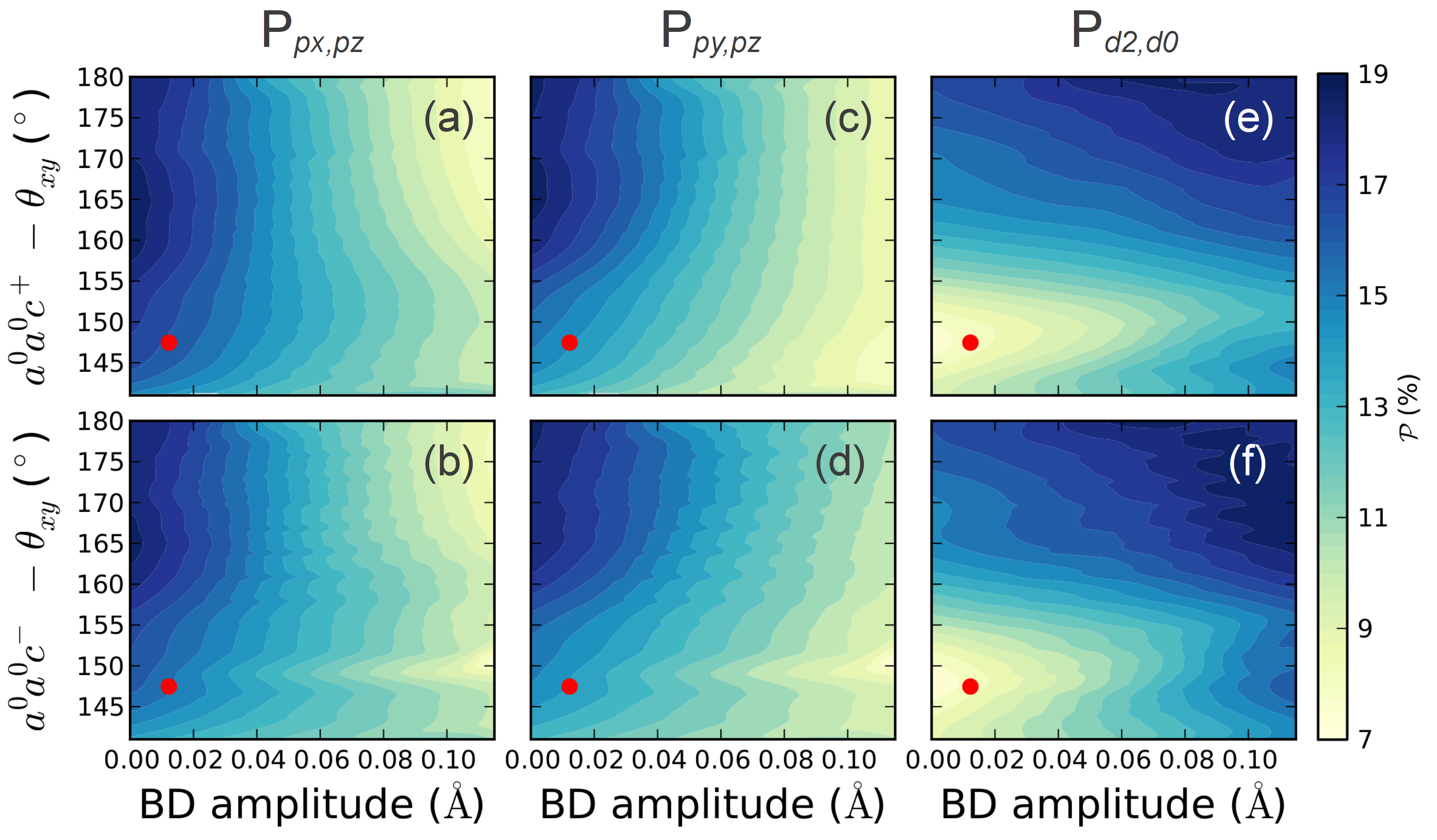}\vspace{-9pt}
  \caption{\label{fig:pol} Orbital polarizations for the apical oxygen atoms $\mathcal{P}_{px,pz}$ (a) and (b), $\mathcal{P}_{py,pz}$ (c) and (d), and Fe(1) site $\mathcal{P}_{d2,d0}$ (e) and (f) for breathing distortions (BD)  coupled with in-phase (upper panels) and out-of-phase (lower panels) rotation patterns, respectively. $a^0a^0c^-$  rotations induce charge transfer to the $p_z$ orbital aligned along the Fe--O apical bond axis; Fe $e_g$ polarization is weakly affected by rotation sense. Filled (red) circles indicate the BD amplitude and the rotation angles for the structures whose DOS are given in \autoref{fig:dos}(b) and (c).
}
\end{figure*}

We now track the change in the orbital polarization, $\mathcal{P}$, as we increase the 
breathing distortion  together with the rotation amplitude of the oxygen 
displacement patterns associated with either the $a^0a^0c^+$ or $a^0a^0c^-$ 
rotations [\autoref{fig:scheme}(a) and \autoref{fig:scheme}(b)] about the $z$-direction.
To compute $\mathcal{P}$, occupancies are calculated by integrating the angular momentum resolved DOS for particular $\left|l m_{l}\right\rangle$, 
obtained by projecting the Kohn-Sham wave functions onto the spherical 
harmonics, starting from the lower edge of the anti-bonding valence band states
up to $E_F$. 
The orbital polarization for the Fe $e_g$ states ($l=2$), \emph{e.g.}, $\mathcal{P}_{d2,d0}$ which specifies the filling of the $d_{x^2-y^2}$ orbital ($m_l=2$) relative to the $d_{3z^2-r^2}$ orbital ($m_l=0$) is 
$\mathcal{P}_{d2,d0}=(n_{x^2-y^2}-n_{3z^2-r^2})/(n_{x^2-y^2}+n_{3z^2-r^2}),$ 
where $n_{x^2-y^2}$ and $n_{3z^2-r^2}$ are the $d_{x^2-y^2}$ and $d_{3z^2-r^2}$ orbital occupancies, respectively.
We first analyze the orbital polarization of the apical [\autoref{fig:pol}(a-d)] and equatorial oxygen atoms in \ce{CaFeO3} (see \autoref{fig:rot_ang}) for all possible 2$p$ orbital pairs with increasing amplitude of rotation and BD under a fixed volume constraint: $\mathcal{P}_{px,py}$, $\mathcal{P}_{px,pz}$ and $\mathcal{P}_{py,pz}$ ($m_l =-1, 0, 1$). 
Next, we evaluate $\mathcal{P}_{d2,d0}$[Fe(1)] and $\mathcal{P}_{d2,d0}$[Fe(2)], \emph{i.e.}, the orbital polarization of each unique Fe site.

The $\mathcal{P}_{px,py}$ polarization of the \emph{apical} oxygen atoms evolves identically for both in-phase and out-of-phase rotations (data not shown); the small polarization value ($0.1\%$) also indicates that both $p_x$ and $p_y$ orbitals are equally populated, consistent with the tetragonal constraints imposed by the Bravais lattice.
Over the structure-distortion space examined,\footnote{%
\autoref{fig:pol} is generated from 345 structures 
and performing self-consistent total energy calculations; linear interpolation of the 
computed orbital polarizations yields the full rotation--BD orbital polarization 
surfaces. We apply a two-dimensional moving-average filter to smooth the data for 
clarity.
} 
 we find the 
$\mathcal{P}_{px,pz}$ and $\mathcal{P}_{py,pz}$ polarization is 
positive with $p_x$ and $p_y$ preferentially occupied over the orbital directed along $z$ ($p_z$). 
Both polarizations are $\approx10\%$ and the 
width of variation is $4\%$. 
These axial polarizations diminish for both rotation flavors if the 
BD increases. 
We understand this behavior as follows: 
The $p_z$ orbital pointing along Fe--O bond axis increases its 
electron occupancy owing to a charge transfer from the equatorial plane, 
containing $p_x$ and $p_y$ orbitals, towards the Fe--O--Fe bond axis [\autoref{fig:scheme}(c)].

The orbital polarization involving $p_z$ is nearly constant with 
respect to increasing \emph{in-phase} rotation and turns out to have a 
minimum at large \emph{out-of-phase} angles and large breathing distortion 
values. 
At fixed BD amplitude, large out-of-phase angles direct the charge transfer 
towards the Fe--O bond along the (rotation) $z$-axis. 
For the \emph{equatorial} oxygen atoms, $\mathcal{P}_{px,py}$ is 
also small ($0.1\%$), indicating the orbitals are equally populated (data not shown). 
$\mathcal{P}_{px,pz}$ and $\mathcal{P}_{py,pz}$ have smaller values 
($1\%$) compared to the apical oxygen atoms in 
all distortion ranges examined. 
For large, physically unreasonable distortions, the $p_z$ orbital not participating in 
the \emph{in-plane} Fe--O bond eventually becomes populated.
Note that oxygen orbital polarizations involving $p_z$ are not affected by small breathing 
distortion values; however, coupling a large breathing amplitude with large rotation angles 
makes $\mathcal{P}$ maximum. 

We now examine the Fe $3d$-orbitals with $e_g$ symmetry. Fe(1) orbital polarization $\mathcal{P}_{d2,d0}$[Fe(1)]  increases with breathing distortion [\autoref{fig:pol}(e-f)], \emph{i.e.}, charges flow from the 
\emph{apical} to \emph{equatorial} bond axes.
With decreasing amplitude of the rotation angle $\theta_{xy}$, the $d_{3z^2-r^2}$ becomes 
preferentially occupied [\autoref{fig:scheme}(d)].
For Fe(2) the $\mathcal{P}_{d2,d0}$[Fe(2)] decreases as the charge flow reverses from the equatorial to apical bond. Both polarizations are minimal for small rotation angle values (further from cubic symmetry), yet the differential occupation appears to be weakly dependent on the rotation sense. The out-of-phase rotations, however, influence the $e_g$ orbitals more than the in-phase ones.
To summarize to here, our analysis reveals that the apical oxygen atoms 
lying along the rotation axis are mainly sensitive to the octahedral rotation 
sense, with \emph{out-of-phase} rotations favoring charge flow to the 
Fe--O \emph{apical} bond. 
The orbital occupancy of the \emph{equatorial} oxygen atoms is weakly 
affected by the rotation pattern sense.\footnote{When only the breathing distortion is present, the compound becomes insulating at high BD amplitudes ($\geq 0.13$ \AA{}). If the BD is absent, rotations alone do not induce a metal-insulator transition  \cite{CRprb12_195144}. At fixed BD, however, there are some dependencies of the electronic band gap on the rotation sense.} 
Moreover, the iron sites are largely sensitive to the breathing distortion, whereby the 
more expanded an octahedron becomes, the greater is the charge transfer 
to the apical bond axes. This effect increases the orbital polarization of 
the corresponding Fe cations; it can also be more readily achieved with the 
$a^0a^0c^-$ rather than the $a^0a^0c^+$ rotation pattern.
These results suggest that sense of the octahedral rotation mainly affects  
the oxygen atoms lying along the rotation axis that ``screen'' the 
metal cations, which are unable to ``see'' the rotation sense effects.

\subsection{Orbital Polarization with Multiple Rotations}

In the experimental CaFeO$_3$ structure, the BD is present together with the 
$a^-a^-c^+$ rotation pattern, where both rotation senses are present and 
out-of-phase rotations are found about two Cartesian directions. 
We now analyze the orbital polarization in CFO structures where the BD is 
coupled with the $a^-a^-c^0$ rotation 
to evaluate the conclusions about the different 
rotation sense discussed above in the presence of multiple rotation axes.
We find polarizations $\mathcal{P}_{px,py}$, $\mathcal{P}_{px,pz}$ and 
$\mathcal{P}_{py,pz}$ of the apical oxygen atoms increase when the rotation angle 
$\theta_{z}$ approaches 180$^\circ$ (data not shown).\footnote{Identical conclusions are obtained from considering $\theta_{xy}$ instead of $\theta_{z}$, \emph{i.e.}, when the breathing distortion is coupled with $a^-a^-c^0$ rotations, $\theta_{xy}$ is an increasing monotonic function of $\theta_{z}$.}
As before, the out-of-phase rotations will transfer charge from the orbitals lying in the plane normal to the rotation axes to the orbital lying along the axes. 
In this case, the rotations in the $xy$-plane induce a flow of 
charge from $p_z$ to both the $p_x$ and $p_y$ orbitals, \emph{i.e.} 
from the Fe--O apical bond into the plane orthogonal to it 
and containing the \emph{apical} oxygen atom and its $p_x$ and $p_y$ orbitals. 
On the other hand, the \emph{equatorial} oxygen atoms experience charge flow 
to their corresponding Fe--O bonds. 
Concerning the two Fe sites, decreasing the $\theta_{z}$ rotation angle moves charge 
from the apical ($d_{3z^2-r^2}$ orbital) to the equatorial ($d_{x^2-y^2}$) axis.
Here, the BD has the  role of enhancing the charge transfer 
induced by the $a^-a^-c^+$ rotation pattern. This last result, together 
with the analysis of the single in-phase and out-of-phase case discussed above, 
reveals that \emph{the out-of-phase rotation pattern favors a 
flow of charge towards its corresponding rotation axis}. 
The rotation ``sense'' becomes an important structural feature over 
the orbital population.

In CaFeO$_3$, the Fe--O equatorial bonds are known to play a fundamental role in the metal-insulator transition \cite{CRprb12_195144}: a change in the equatorial bond-lengths together with the octahedral rotations affects the $d_{x^2-y^2}$/($p_x$,$p_y$) orbital hybridization responsible for the formation of a highly covalent electronic configuration, gapping the system.
An alternative route for tuning the band gap emerges from our results. If the Fe--O bond lengths are not altered, then the $a^-a^-c^0$ out-of-phase rotation pattern controls the amount of charge filling the equatorial orbitals, hence the degree of hybridization, which determines the stability of the insulating phase.

\begin{figure}[b]
\centering
\includegraphics[width=0.49\textwidth,clip]{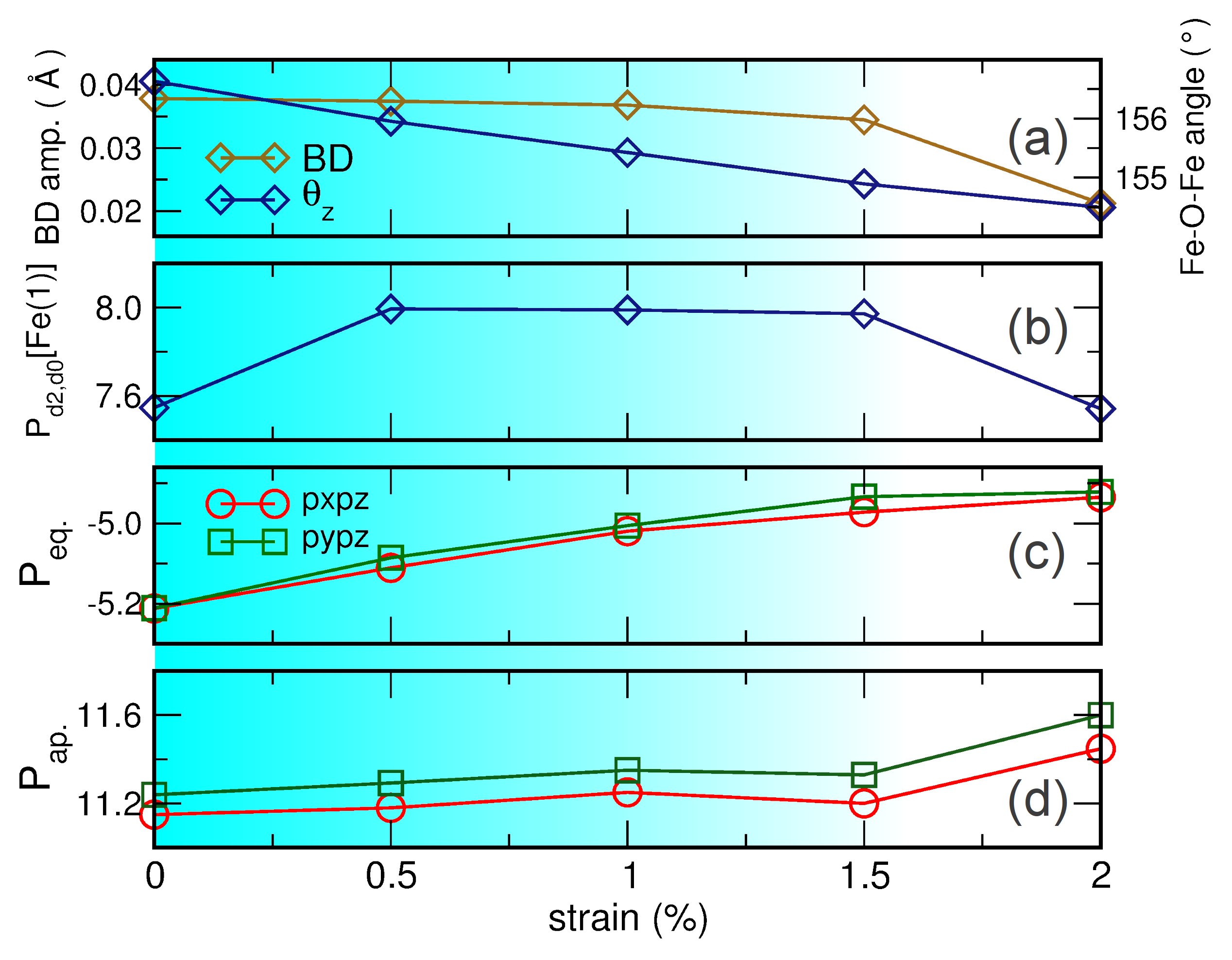}\vspace{-10pt}
\caption{\label{fig:strain} BD amplitudes and $\theta_z$ angles (a), orbital polarization (\%)  of (b) Fe(1) site, (c) equatorial (eq.) and (d) apical (ap.) oxygen atoms for relaxed CaFeO$_3$ structures under biaxial strain. 
The shaded area indicates strain values for insulating structures; the MIT occurs between 1-1.5\% strain.}
\end{figure}

\subsection{Strain--rotation and Orbital Polarization Coupling}

Strain can enhance the octahedral rotations in perovskites \cite{Rondinelli/May/Freeland:2012}, 
which in turn alter the band structure \cite{BFNprl11_146804,Rondinelli/Coh:2011,PhysRevLett.106.237601,LLWHLLMapl11_202110}.
Tensile bi-axial strain promotes the $a^-a^-c^+$ rotation pattern in $c$-oriented orthorhombic perovskites thin films \cite{RSav11_3363}. 
To explore the effect of  strain on rotation-induced $\mathcal{P}$, we compute the 
equilibrium structure of CaFeO$_3$ under biaxial tensile strain, relaxing ionic positions and the $c$-axis length. 

First, we evaluate the effect of strain on the atomic structure.
We find that $\theta_z$ deviates more from 180$^\circ$ with 
increasing tensile strain [\autoref{fig:strain}(a)]. 
Consistent with our previous observation, when the BD and 
$a^-a^-c^0$ rotations are active, the orbital polarization is 
enhanced along the Fe--O equatorial bonds with greater tensile strain.
Importantly, we note that within the insulating phase as strain is increased, the BD largely remains constant while the octahedral rotations are enhanced to accommodate the strain energy [\autoref{fig:strain}(a)].
The result of the decreased $\theta_z$ rotation angle is further charge localization in the $d_{x^2-y^2}$ [\autoref{fig:strain}(b)] and $p_x$ and $p_y$ orbitals of equatorial oxygen atoms [\autoref{fig:strain}(c)], hence about the equatorial Fe--O bonds supporting 
the insulating state and leaving unaltered the polarization of apical oxygen atoms [\autoref{fig:strain}(d)].
Our results show that the O $2p$ and Fe $e_g$ orbital polarizations are highly sensitive to the strain 
state and are mediated largely by the change in the out-of-phase octahedral rotations ($\theta_z$).
The tensile strain, enhances the $a^-a^-c^0$ rotation pattern, induces a charge transfer into those orbitals responsible for the gap
(\emph{i.e.} $d_{x^2-y^2}$, and $p_x$/$p_y$ of equatorial oxygen atoms): beyond a critical value, the charge becomes delocalized and the insulating gap closes to form a metallic state.
We observe that the insulating phase is lost for strain values greater than 1\%, which is in contrast to 
that seen in rare-earth nickelates \cite{PhysRevB.85.214431}, suggesting the rotation--orbital polarization 
interaction and the BD may be either complementary or antagonistic to the MIT.

The out-of-phase rotations are also responsible for the broadening of the $p_x$ and $p_y$ orbitals about $E_F$, which we conjecture would reduce the sharpness of the first-order phase transition. 
Note that the critical transition temperature for the MIT is also believed to be 
correlated with the size of the BD, our results therefore suggest that strain-engineering of the MIT temperature could be challenging in ferrates.
On the other hand, the structural transition temperature can be kept fixed 
with increasing strain while the electronic transition can still be tuned by means of the octahedral rotation-induced orbital polarizations.
\section{Conclusions}

In summary, we have shown how the sense of the octahedral rotation pattern and 
not only the magnitude of the rotation distortion is important in engineering 
the low energy electronic band structure. 
We find that while the \emph{in-phase} rotations weakly affect the local electronic distribution, the \emph{out-of-phase}  rotation pattern directs a flow of charge towards its corresponding rotation axis.
The size of the rotation-sense orbital polarization we compute should be experimentally detectable with orbital reflectometry techniques \cite{BHBGFYACHBZWKHKnm11_189},  across the orthorhombic ($a^-a^-c^+$) to monoclinic  ($a^-b^-c^+$) phase transition of CaFeO$_3$ near room temperature. Here, the out-of-phase rotation pattern is no longer uniform along [110].
For charge transfer perovskite oxides, control over the phase of the rotations 
via heteroepitaxial strain, substrate proximity effects, or 
superlattice formation provides a promising alternative route to tailor the local orbital polarizations, hence electronic transitions, in 
correlated materials near electronic phase boundaries. 
Due to the complexity of these interactions driven by cation substitution, we hope this work motivates further experimental studies to evaluate the  mechanisms proposed here, and the degree to which they compete or cooperate across  a MIT.

\begin{acknowledgments}
A.C.\ and J.M.R.\ were supported by 
ONR under grant no.\ N00014-11-1-0664.
This work used the Extreme Science and Engineering Discovery Environment (XSEDE), which is supported by NSF (OCI-1053575). The use of VESTA software is also acknowledged \cite{MIjac08_653}.
\end{acknowledgments}


\begin{thebibliography}{45}%
\makeatletter
\providecommand \@ifxundefined [1]{%
 \@ifx{#1\undefined}
}%
\providecommand \@ifnum [1]{%
 \ifnum #1\expandafter \@firstoftwo
 \else \expandafter \@secondoftwo
 \fi
}%
\providecommand \@ifx [1]{%
 \ifx #1\expandafter \@firstoftwo
 \else \expandafter \@secondoftwo
 \fi
}%
\providecommand \natexlab [1]{#1}%
\providecommand \enquote  [1]{``#1''}%
\providecommand \bibnamefont  [1]{#1}%
\providecommand \bibfnamefont [1]{#1}%
\providecommand \citenamefont [1]{#1}%
\providecommand \href@noop [0]{\@secondoftwo}%
\providecommand \href [0]{\begingroup \@sanitize@url \@href}%
\providecommand \@href[1]{\@@startlink{#1}\@@href}%
\providecommand \@@href[1]{\endgroup#1\@@endlink}%
\providecommand \@sanitize@url [0]{\catcode `\\12\catcode `\$12\catcode
  `\&12\catcode `\#12\catcode `\^12\catcode `\_12\catcode `\%12\relax}%
\providecommand \@@startlink[1]{}%
\providecommand \@@endlink[0]{}%
\providecommand \url  [0]{\begingroup\@sanitize@url \@url }%
\providecommand \@url [1]{\endgroup\@href {#1}{\urlprefix }}%
\providecommand \urlprefix  [0]{URL }%
\providecommand \Eprint [0]{\href }%
\providecommand \doibase [0]{http://dx.doi.org/}%
\providecommand \selectlanguage [0]{\@gobble}%
\providecommand \bibinfo  [0]{\@secondoftwo}%
\providecommand \bibfield  [0]{\@secondoftwo}%
\providecommand \translation [1]{[#1]}%
\providecommand \BibitemOpen [0]{}%
\providecommand \bibitemStop [0]{}%
\providecommand \bibitemNoStop [0]{.\EOS\space}%
\providecommand \EOS [0]{\spacefactor3000\relax}%
\providecommand \BibitemShut  [1]{\csname bibitem#1\endcsname}%
\let\auto@bib@innerbib\@empty
\bibitem [{\citenamefont {Dagotto}(2005)}]{Dagotto:2005}%
  \BibitemOpen
  \bibfield  {author} {\bibinfo {author} {\bibfnamefont {E.}~\bibnamefont
  {Dagotto}},\ }\href {\doibase 10.1126/science.1107559} {\bibfield  {journal}
  {\bibinfo  {journal} {Science}\ }\textbf {\bibinfo {volume} {309}},\ \bibinfo
  {pages} {257} (\bibinfo {year} {2005})}\BibitemShut {NoStop}%
\bibitem [{\citenamefont {Imada}\ \emph {et~al.}(1998)\citenamefont {Imada},
  \citenamefont {Fujimori},\ and\ \citenamefont
  {Tokura}}]{Imada/Fujimori/Tokura:1998}%
  \BibitemOpen
  \bibfield  {author} {\bibinfo {author} {\bibfnamefont {M.}~\bibnamefont
  {Imada}}, \bibinfo {author} {\bibfnamefont {A.}~\bibnamefont {Fujimori}}, \
  and\ \bibinfo {author} {\bibfnamefont {Y.}~\bibnamefont {Tokura}},\ }\href
  {\doibase 10.1103/RevModPhys.70.1039} {\bibfield  {journal} {\bibinfo
  {journal} {Rev. Mod. Phys.}\ }\textbf {\bibinfo {volume} {70}},\ \bibinfo
  {pages} {1039} (\bibinfo {year} {1998})}\BibitemShut {NoStop}%
\bibitem [{\citenamefont {Tokura}\ and\ \citenamefont
  {Nagaosa}(2000)}]{TNs00_462}%
  \BibitemOpen
  \bibfield  {author} {\bibinfo {author} {\bibfnamefont {Y.}~\bibnamefont
  {Tokura}}\ and\ \bibinfo {author} {\bibfnamefont {N.}~\bibnamefont
  {Nagaosa}},\ }\href {\doibase 10.1126/science.288.5465.462} {\bibfield
  {journal} {\bibinfo  {journal} {Science}\ }\textbf {\bibinfo {volume}
  {288}},\ \bibinfo {pages} {462} (\bibinfo {year} {2000})}\BibitemShut
  {NoStop}%
\bibitem [{\citenamefont {Salamon}\ and\ \citenamefont
  {Jaime}(2001)}]{Salamon/Marcelo:2001}%
  \BibitemOpen
  \bibfield  {author} {\bibinfo {author} {\bibfnamefont {M.~B.}\ \bibnamefont
  {Salamon}}\ and\ \bibinfo {author} {\bibfnamefont {M.}~\bibnamefont
  {Jaime}},\ }\href {\doibase 10.1103/RevModPhys.73.583} {\bibfield  {journal}
  {\bibinfo  {journal} {Rev. Mod. Phys.}\ }\textbf {\bibinfo {volume} {73}},\
  \bibinfo {pages} {583} (\bibinfo {year} {2001})}\BibitemShut {NoStop}%
\bibitem [{\citenamefont {Werner}\ and\ \citenamefont
  {Millis}(2007)}]{Werner/Millis:2007}%
  \BibitemOpen
  \bibfield  {author} {\bibinfo {author} {\bibfnamefont {P.}~\bibnamefont
  {Werner}}\ and\ \bibinfo {author} {\bibfnamefont {A.~J.}\ \bibnamefont
  {Millis}},\ }\href {\doibase 10.1103/PhysRevLett.99.126405} {\bibfield
  {journal} {\bibinfo  {journal} {Phys. Rev. Lett.}\ }\textbf {\bibinfo
  {volume} {99}},\ \bibinfo {pages} {126405} (\bibinfo {year}
  {2007})}\BibitemShut {NoStop}%
\bibitem [{\citenamefont {Freeland}\ \emph {et~al.}(2011)\citenamefont
  {Freeland}, \citenamefont {Liu}, \citenamefont {Kareev}, \citenamefont
  {Gray}, \citenamefont {Kim}, \citenamefont {Ryan}, \citenamefont
  {Pentcheva},\ and\ \citenamefont {Chakhalian}}]{0295-5075-96-5-57004}%
  \BibitemOpen
  \bibfield  {author} {\bibinfo {author} {\bibfnamefont {J.~W.}\ \bibnamefont
  {Freeland}}, \bibinfo {author} {\bibfnamefont {J.}~\bibnamefont {Liu}},
  \bibinfo {author} {\bibfnamefont {M.}~\bibnamefont {Kareev}}, \bibinfo
  {author} {\bibfnamefont {B.}~\bibnamefont {Gray}}, \bibinfo {author}
  {\bibfnamefont {J.~W.}\ \bibnamefont {Kim}}, \bibinfo {author} {\bibfnamefont
  {P.}~\bibnamefont {Ryan}}, \bibinfo {author} {\bibfnamefont {R.}~\bibnamefont
  {Pentcheva}}, \ and\ \bibinfo {author} {\bibfnamefont {J.}~\bibnamefont
  {Chakhalian}},\ }\href {http://stacks.iop.org/0295-5075/96/i=5/a=57004}
  {\bibfield  {journal} {\bibinfo  {journal} {Europhys.\ Lett.}\ }\textbf
  {\bibinfo {volume} {96}},\ \bibinfo {pages} {57004} (\bibinfo {year}
  {2011})}\BibitemShut {NoStop}%
\bibitem [{\citenamefont {Moon}\ \emph {et~al.}(2012)\citenamefont {Moon},
  \citenamefont {Rondinelli}, \citenamefont {Prasai}, \citenamefont {Gray},
  \citenamefont {Kareev}, \citenamefont {Chakhalian},\ and\ \citenamefont
  {Cohn}}]{Moon/Chakhalian/Rondinelli:2012}%
  \BibitemOpen
  \bibfield  {author} {\bibinfo {author} {\bibfnamefont {E.~J.}\ \bibnamefont
  {Moon}}, \bibinfo {author} {\bibfnamefont {J.~M.}\ \bibnamefont
  {Rondinelli}}, \bibinfo {author} {\bibfnamefont {N.}~\bibnamefont {Prasai}},
  \bibinfo {author} {\bibfnamefont {B.~A.}\ \bibnamefont {Gray}}, \bibinfo
  {author} {\bibfnamefont {M.}~\bibnamefont {Kareev}}, \bibinfo {author}
  {\bibfnamefont {J.}~\bibnamefont {Chakhalian}}, \ and\ \bibinfo {author}
  {\bibfnamefont {J.~L.}\ \bibnamefont {Cohn}},\ }\href {\doibase
  10.1103/PhysRevB.85.121106} {\bibfield  {journal} {\bibinfo  {journal}
  {Phys.\ Rev.\ B}\ }\textbf {\bibinfo {volume} {85}},\ \bibinfo {pages}
  {121106} (\bibinfo {year} {2012})}\BibitemShut {NoStop}%
\bibitem [{\citenamefont {Han}\ \emph {et~al.}(2010)\citenamefont {Han},
  \citenamefont {Marianetti},\ and\ \citenamefont
  {Millis}}]{PhysRevB.82.134408}%
  \BibitemOpen
  \bibfield  {author} {\bibinfo {author} {\bibfnamefont {M.~J.}\ \bibnamefont
  {Han}}, \bibinfo {author} {\bibfnamefont {C.~A.}\ \bibnamefont {Marianetti}},
  \ and\ \bibinfo {author} {\bibfnamefont {A.~J.}\ \bibnamefont {Millis}},\
  }\href {\doibase 10.1103/PhysRevB.82.134408} {\bibfield  {journal} {\bibinfo
  {journal} {Phys. Rev. B}\ }\textbf {\bibinfo {volume} {82}},\ \bibinfo
  {pages} {134408} (\bibinfo {year} {2010})}\BibitemShut {NoStop}%
\bibitem [{\citenamefont {Liu}\ \emph {et~al.}(2011)\citenamefont {Liu},
  \citenamefont {Okamoto}, \citenamefont {van Veenendaal}, \citenamefont
  {Kareev}, \citenamefont {Gray}, \citenamefont {Ryan}, \citenamefont
  {Freeland},\ and\ \citenamefont {Chakhalian}}]{PhysRevB.83.161102}%
  \BibitemOpen
  \bibfield  {author} {\bibinfo {author} {\bibfnamefont {J.}~\bibnamefont
  {Liu}}, \bibinfo {author} {\bibfnamefont {S.}~\bibnamefont {Okamoto}},
  \bibinfo {author} {\bibfnamefont {M.}~\bibnamefont {van Veenendaal}},
  \bibinfo {author} {\bibfnamefont {M.}~\bibnamefont {Kareev}}, \bibinfo
  {author} {\bibfnamefont {B.}~\bibnamefont {Gray}}, \bibinfo {author}
  {\bibfnamefont {P.}~\bibnamefont {Ryan}}, \bibinfo {author} {\bibfnamefont
  {J.~W.}\ \bibnamefont {Freeland}}, \ and\ \bibinfo {author} {\bibfnamefont
  {J.}~\bibnamefont {Chakhalian}},\ }\href {\doibase
  10.1103/PhysRevB.83.161102} {\bibfield  {journal} {\bibinfo  {journal} {Phys.
  Rev. B}\ }\textbf {\bibinfo {volume} {83}},\ \bibinfo {pages} {161102}
  (\bibinfo {year} {2011})}\BibitemShut {NoStop}%
\bibitem [{\citenamefont {Hansmann}\ \emph {et~al.}(2010)\citenamefont
  {Hansmann}, \citenamefont {Toschi}, \citenamefont {Yang}, \citenamefont
  {Andersen},\ and\ \citenamefont {Held}}]{PhysRevB.82.235123}%
  \BibitemOpen
  \bibfield  {author} {\bibinfo {author} {\bibfnamefont {P.}~\bibnamefont
  {Hansmann}}, \bibinfo {author} {\bibfnamefont {A.}~\bibnamefont {Toschi}},
  \bibinfo {author} {\bibfnamefont {X.}~\bibnamefont {Yang}}, \bibinfo {author}
  {\bibfnamefont {O.~K.}\ \bibnamefont {Andersen}}, \ and\ \bibinfo {author}
  {\bibfnamefont {K.}~\bibnamefont {Held}},\ }\href {\doibase
  10.1103/PhysRevB.82.235123} {\bibfield  {journal} {\bibinfo  {journal} {Phys.
  Rev. B}\ }\textbf {\bibinfo {volume} {82}},\ \bibinfo {pages} {235123}
  (\bibinfo {year} {2010})}\BibitemShut {NoStop}%
\bibitem [{\citenamefont {Burdett}(1975)}]{doi:10.1021/ic50146a042}%
  \BibitemOpen
  \bibfield  {author} {\bibinfo {author} {\bibfnamefont {J.~K.}\ \bibnamefont
  {Burdett}},\ }\href {\doibase 10.1021/ic50146a042} {\bibfield  {journal}
  {\bibinfo  {journal} {Inorg.\ Chem.}\ }\textbf {\bibinfo {volume} {14}},\
  \bibinfo {pages} {931} (\bibinfo {year} {1975})}\BibitemShut {NoStop}%
\bibitem [{\citenamefont {Liu}(2011)}]{Lprb11_235136}%
  \BibitemOpen
  \bibfield  {author} {\bibinfo {author} {\bibfnamefont {G.-Q.}\ \bibnamefont
  {Liu}},\ }\href {\doibase 10.1103/PhysRevB.84.235136} {\bibfield  {journal}
  {\bibinfo  {journal} {Phys. Rev. B}\ }\textbf {\bibinfo {volume} {84}},\
  \bibinfo {pages} {235136} (\bibinfo {year} {2011})}\BibitemShut {NoStop}%
\bibitem [{\citenamefont {Ha}\ and\ \citenamefont
  {Ramanathan}(2011)}]{ha:071101}%
  \BibitemOpen
  \bibfield  {author} {\bibinfo {author} {\bibfnamefont {S.~D.}\ \bibnamefont
  {Ha}}\ and\ \bibinfo {author} {\bibfnamefont {S.}~\bibnamefont
  {Ramanathan}},\ }\href {\doibase 10.1063/1.3640806} {\bibfield  {journal}
  {\bibinfo  {journal} {J.\ App.\ Phys.}\ }\textbf {\bibinfo {volume} {110}},\
  \bibinfo {eid} {071101} (\bibinfo {year} {2011})}\BibitemShut {NoStop}%
\bibitem [{\citenamefont {Tomioka}\ \emph {et~al.}(2001)\citenamefont
  {Tomioka}, \citenamefont {Okuda}, \citenamefont {Okimoto}, \citenamefont
  {Asamitsu}, \citenamefont {Kuwahara},\ and\ \citenamefont
  {Tokura}}]{TOOAKTjac01_27}%
  \BibitemOpen
  \bibfield  {author} {\bibinfo {author} {\bibfnamefont {Y.}~\bibnamefont
  {Tomioka}}, \bibinfo {author} {\bibfnamefont {T.}~\bibnamefont {Okuda}},
  \bibinfo {author} {\bibfnamefont {Y.}~\bibnamefont {Okimoto}}, \bibinfo
  {author} {\bibfnamefont {A.}~\bibnamefont {Asamitsu}}, \bibinfo {author}
  {\bibfnamefont {H.}~\bibnamefont {Kuwahara}}, \ and\ \bibinfo {author}
  {\bibfnamefont {Y.}~\bibnamefont {Tokura}},\ }\href {\doibase
  http://dx.doi.org/10.1016/S0925-8388(01)01222-1} {\bibfield  {journal}
  {\bibinfo  {journal} {J. All. Comp.}\ }\textbf {\bibinfo {volume} {326}},\
  \bibinfo {pages} {27} (\bibinfo {year} {2001})}\BibitemShut {NoStop}%
\bibitem [{\citenamefont {Whangbo}\ \emph {et~al.}(2002)\citenamefont
  {Whangbo}, \citenamefont {Koo}, \citenamefont {Villesuzanne},\ and\
  \citenamefont {Pouchard}}]{WKVPic02_1920}%
  \BibitemOpen
  \bibfield  {author} {\bibinfo {author} {\bibfnamefont {M.~H.}\ \bibnamefont
  {Whangbo}}, \bibinfo {author} {\bibfnamefont {H.~J.}\ \bibnamefont {Koo}},
  \bibinfo {author} {\bibfnamefont {A.}~\bibnamefont {Villesuzanne}}, \ and\
  \bibinfo {author} {\bibfnamefont {M.}~\bibnamefont {Pouchard}},\ }\href
  {http://pubs.acs.org/doi/abs/10.1021/ic0110427} {\bibfield  {journal}
  {\bibinfo  {journal} {Inorg. Chem.}\ }\textbf {\bibinfo {volume} {41}},\
  \bibinfo {pages} {1920} (\bibinfo {year} {2002})}\BibitemShut {NoStop}%
\bibitem [{\citenamefont {Park}\ \emph {et~al.}(1999)\citenamefont {Park},
  \citenamefont {Ishikawa}, \citenamefont {Tokura}, \citenamefont {Li},\ and\
  \citenamefont {Matsui}}]{PITLMprb99_10788}%
  \BibitemOpen
  \bibfield  {author} {\bibinfo {author} {\bibfnamefont {S.~K.}\ \bibnamefont
  {Park}}, \bibinfo {author} {\bibfnamefont {T.}~\bibnamefont {Ishikawa}},
  \bibinfo {author} {\bibfnamefont {Y.}~\bibnamefont {Tokura}}, \bibinfo
  {author} {\bibfnamefont {J.~Q.}\ \bibnamefont {Li}}, \ and\ \bibinfo {author}
  {\bibfnamefont {Y.}~\bibnamefont {Matsui}},\ }\href
  {http://prb.aps.org/abstract/PRB/v60/i15/p10788_1} {\bibfield  {journal}
  {\bibinfo  {journal} {Phys. Rev. B}\ }\textbf {\bibinfo {volume} {60}},\
  \bibinfo {pages} {10788} (\bibinfo {year} {1999})}\BibitemShut {NoStop}%
\bibitem [{\citenamefont {Mostovoy}(2005{\natexlab{a}})}]{Mjpcm05_753}%
  \BibitemOpen
  \bibfield  {author} {\bibinfo {author} {\bibfnamefont {M.}~\bibnamefont
  {Mostovoy}},\ }\href@noop {} {\bibfield  {journal} {\bibinfo  {journal} {J.
  Phys.: Condens. Matter}\ }\textbf {\bibinfo {volume} {17}},\ \bibinfo {pages}
  {S753} (\bibinfo {year} {2005}{\natexlab{a}})}\BibitemShut {NoStop}%
\bibitem [{\citenamefont {Mostovoy}(2005{\natexlab{b}})}]{Mprl05_137205}%
  \BibitemOpen
  \bibfield  {author} {\bibinfo {author} {\bibfnamefont {M.}~\bibnamefont
  {Mostovoy}},\ }\href {\doibase 10.1103/PhysRevLett.94.137205} {\bibfield
  {journal} {\bibinfo  {journal} {Phys. Rev. Letters}\ }\textbf {\bibinfo
  {volume} {94}},\ \bibinfo {pages} {137205} (\bibinfo {year}
  {2005}{\natexlab{b}})}\BibitemShut {NoStop}%
\bibitem [{\citenamefont {Saha-Dasgupta}\ \emph {et~al.}(2005)\citenamefont
  {Saha-Dasgupta}, \citenamefont {Popovi\ifmmode~\acute{c}\else \'{c}\fi{}},\
  and\ \citenamefont {Satpathy}}]{SPSprb05_45143}%
  \BibitemOpen
  \bibfield  {author} {\bibinfo {author} {\bibfnamefont {T.}~\bibnamefont
  {Saha-Dasgupta}}, \bibinfo {author} {\bibfnamefont {Z.~S.}\ \bibnamefont
  {Popovi\ifmmode~\acute{c}\else \'{c}\fi{}}}, \ and\ \bibinfo {author}
  {\bibfnamefont {S.}~\bibnamefont {Satpathy}},\ }\href {\doibase
  10.1103/PhysRevB.72.045143} {\bibfield  {journal} {\bibinfo  {journal} {Phys.
  Rev. B}\ }\textbf {\bibinfo {volume} {72}},\ \bibinfo {pages} {045143}
  (\bibinfo {year} {2005})}\BibitemShut {NoStop}%
\bibitem [{\citenamefont {Cammarata}\ and\ \citenamefont
  {Rondinelli}(2012)}]{CRprb12_195144}%
  \BibitemOpen
  \bibfield  {author} {\bibinfo {author} {\bibfnamefont {A.}~\bibnamefont
  {Cammarata}}\ and\ \bibinfo {author} {\bibfnamefont {J.~M.}\ \bibnamefont
  {Rondinelli}},\ }\href {\doibase 10.1103/PhysRevB.86.195144} {\bibfield
  {journal} {\bibinfo  {journal} {Phys. Rev. B}\ }\textbf {\bibinfo {volume}
  {86}},\ \bibinfo {pages} {195144} (\bibinfo {year} {2012})}\BibitemShut
  {NoStop}%
\bibitem [{\citenamefont {Lee}\ \emph {et~al.}(2011)\citenamefont {Lee},
  \citenamefont {Chen},\ and\ \citenamefont {Balents}}]{PhysRevB.84.165119}%
  \BibitemOpen
  \bibfield  {author} {\bibinfo {author} {\bibfnamefont {S.}~\bibnamefont
  {Lee}}, \bibinfo {author} {\bibfnamefont {R.}~\bibnamefont {Chen}}, \ and\
  \bibinfo {author} {\bibfnamefont {L.}~\bibnamefont {Balents}},\ }\href
  {\doibase 10.1103/PhysRevB.84.165119} {\bibfield  {journal} {\bibinfo
  {journal} {Phys. Rev. B}\ }\textbf {\bibinfo {volume} {84}},\ \bibinfo
  {pages} {165119} (\bibinfo {year} {2011})}\BibitemShut {NoStop}%
\bibitem [{\citenamefont {Alonso}\ \emph {et~al.}(1999)\citenamefont {Alonso},
  \citenamefont {Garc\'ia-Mu\~noz}, \citenamefont {Fern\'andez-D\'iaz},
  \citenamefont {Aranda}, \citenamefont {Mart\'inez-Lope},\ and\ \citenamefont
  {Casais}}]{PhysRevLett.82.3871}%
  \BibitemOpen
  \bibfield  {author} {\bibinfo {author} {\bibfnamefont {J.~A.}\ \bibnamefont
  {Alonso}}, \bibinfo {author} {\bibfnamefont {J.~L.}\ \bibnamefont
  {Garc\'ia-Mu\~noz}}, \bibinfo {author} {\bibfnamefont {M.~T.}\ \bibnamefont
  {Fern\'andez-D\'iaz}}, \bibinfo {author} {\bibfnamefont {M.~A.~G.}\
  \bibnamefont {Aranda}}, \bibinfo {author} {\bibfnamefont {M.~J.}\
  \bibnamefont {Mart\'inez-Lope}}, \ and\ \bibinfo {author} {\bibfnamefont
  {M.~T.}\ \bibnamefont {Casais}},\ }\href {\doibase
  10.1103/PhysRevLett.82.3871} {\bibfield  {journal} {\bibinfo  {journal}
  {Phys. Rev. Lett.}\ }\textbf {\bibinfo {volume} {82}},\ \bibinfo {pages}
  {3871} (\bibinfo {year} {1999})}\BibitemShut {NoStop}%
\bibitem [{\citenamefont {Park}\ \emph {et~al.}(2012)\citenamefont {Park},
  \citenamefont {Millis},\ and\ \citenamefont
  {Marianetti}}]{PhysRevLett.109.156402}%
  \BibitemOpen
  \bibfield  {author} {\bibinfo {author} {\bibfnamefont {H.}~\bibnamefont
  {Park}}, \bibinfo {author} {\bibfnamefont {A.~J.}\ \bibnamefont {Millis}}, \
  and\ \bibinfo {author} {\bibfnamefont {C.~A.}\ \bibnamefont {Marianetti}},\
  }\href {\doibase 10.1103/PhysRevLett.109.156402} {\bibfield  {journal}
  {\bibinfo  {journal} {Phys. Rev. Lett.}\ }\textbf {\bibinfo {volume} {109}},\
  \bibinfo {pages} {156402} (\bibinfo {year} {2012})}\BibitemShut {NoStop}%
\bibitem [{\citenamefont {Prosandeev}\ \emph {et~al.}(2012)\citenamefont
  {Prosandeev}, \citenamefont {Bellaiche},\ and\ \citenamefont
  {\'I\~niguez}}]{PhysRevB.85.214431}%
  \BibitemOpen
  \bibfield  {author} {\bibinfo {author} {\bibfnamefont {S.}~\bibnamefont
  {Prosandeev}}, \bibinfo {author} {\bibfnamefont {L.}~\bibnamefont
  {Bellaiche}}, \ and\ \bibinfo {author} {\bibfnamefont {J.}~\bibnamefont
  {\'I\~niguez}},\ }\href {\doibase 10.1103/PhysRevB.85.214431} {\bibfield
  {journal} {\bibinfo  {journal} {Phys. Rev. B}\ }\textbf {\bibinfo {volume}
  {85}},\ \bibinfo {pages} {214431} (\bibinfo {year} {2012})}\BibitemShut
  {NoStop}%
\bibitem [{Note1()}]{Note1}%
  \BibitemOpen
  \bibinfo {note} {Strictly, the low-temperature insulating monoclinic phase
  adopts the $a^-b^-c^+$, and thus the angle of rotations about each axis are
  unique, but these deviations in the bond angles are small.}\BibitemShut
  {Stop}%
\bibitem [{\citenamefont {Glazer}(1972)}]{Glazer:1972}%
  \BibitemOpen
  \bibfield  {author} {\bibinfo {author} {\bibfnamefont {A.~M.}\ \bibnamefont
  {Glazer}},\ }\href {http://scripts.iucr.org/cgi-bin/paper?s0567740872007976}
  {\bibfield  {journal} {\bibinfo  {journal} {Acta Cryst. B}\ }\textbf
  {\bibinfo {volume} {28}},\ \bibinfo {pages} {3384} (\bibinfo {year}
  {1972})}\BibitemShut {NoStop}%
\bibitem [{\citenamefont {Bl\"ochl}(1994)}]{Bprb94_17953}%
  \BibitemOpen
  \bibfield  {author} {\bibinfo {author} {\bibfnamefont {P.~E.}\ \bibnamefont
  {Bl\"ochl}},\ }\href {\doibase 10.1103/PhysRevB.50.17953} {\bibfield
  {journal} {\bibinfo  {journal} {Phys. Rev. B}\ }\textbf {\bibinfo {volume}
  {50}},\ \bibinfo {pages} {17953} (\bibinfo {year} {1994})}\BibitemShut
  {NoStop}%
\bibitem [{\citenamefont {Kresse}\ and\ \citenamefont
  {Furthm\"uller}(1996)}]{KFcms96_15}%
  \BibitemOpen
  \bibfield  {author} {\bibinfo {author} {\bibfnamefont {G.}~\bibnamefont
  {Kresse}}\ and\ \bibinfo {author} {\bibfnamefont {J.}~\bibnamefont
  {Furthm\"uller}},\ }\href {\doibase 10.1016/0927-0256(96)00008-0} {\bibfield
  {journal} {\bibinfo  {journal} {Comput. Mat. Sci.}\ }\textbf {\bibinfo
  {volume} {6}},\ \bibinfo {pages} {15 } (\bibinfo {year} {1996})}\BibitemShut
  {NoStop}%
\bibitem [{\citenamefont {Kresse}\ and\ \citenamefont
  {Joubert}(1999)}]{KJprb99_1758}%
  \BibitemOpen
  \bibfield  {author} {\bibinfo {author} {\bibfnamefont {G.}~\bibnamefont
  {Kresse}}\ and\ \bibinfo {author} {\bibfnamefont {D.}~\bibnamefont
  {Joubert}},\ }\href {\doibase 10.1103/PhysRevB.59.1758} {\bibfield  {journal}
  {\bibinfo  {journal} {Phys. Rev. B}\ }\textbf {\bibinfo {volume} {59}},\
  \bibinfo {pages} {1758} (\bibinfo {year} {1999})}\BibitemShut {NoStop}%
\bibitem [{\citenamefont {Perdew}\ \emph {et~al.}(1996)\citenamefont {Perdew},
  \citenamefont {Burke},\ and\ \citenamefont {Ernzerhof}}]{PBE96_3865}%
  \BibitemOpen
  \bibfield  {author} {\bibinfo {author} {\bibfnamefont {J.~P.}\ \bibnamefont
  {Perdew}}, \bibinfo {author} {\bibfnamefont {K.}~\bibnamefont {Burke}}, \
  and\ \bibinfo {author} {\bibfnamefont {M.}~\bibnamefont {Ernzerhof}},\ }\href
  {\doibase 10.1103/PhysRevLett.77.3865} {\bibfield  {journal} {\bibinfo
  {journal} {Phys. Rev. Lett.}\ }\textbf {\bibinfo {volume} {77}},\ \bibinfo
  {pages} {3865} (\bibinfo {year} {1996})}\BibitemShut {NoStop}%
\bibitem [{\citenamefont {Perdew}\ \emph {et~al.}(2008)\citenamefont {Perdew},
  \citenamefont {Ruzsinszky}, \citenamefont {Csonka}, \citenamefont {Vydrov},
  \citenamefont {Scuseria}, \citenamefont {Constantin}, \citenamefont {Zhou},\
  and\ \citenamefont {Burke}}]{PRCVSCZBprl_08136406}%
  \BibitemOpen
  \bibfield  {author} {\bibinfo {author} {\bibfnamefont {J.~P.}\ \bibnamefont
  {Perdew}}, \bibinfo {author} {\bibfnamefont {A.}~\bibnamefont {Ruzsinszky}},
  \bibinfo {author} {\bibfnamefont {G.~I.}\ \bibnamefont {Csonka}}, \bibinfo
  {author} {\bibfnamefont {O.~A.}\ \bibnamefont {Vydrov}}, \bibinfo {author}
  {\bibfnamefont {G.~E.}\ \bibnamefont {Scuseria}}, \bibinfo {author}
  {\bibfnamefont {L.~A.}\ \bibnamefont {Constantin}}, \bibinfo {author}
  {\bibfnamefont {X.}~\bibnamefont {Zhou}}, \ and\ \bibinfo {author}
  {\bibfnamefont {K.}~\bibnamefont {Burke}},\ }\href {\doibase
  10.1103/PhysRevLett.100.136406} {\bibfield  {journal} {\bibinfo  {journal}
  {Phys. Rev. Lett.}\ }\textbf {\bibinfo {volume} {100}},\ \bibinfo {pages}
  {136406} (\bibinfo {year} {2008})}\BibitemShut {NoStop}%
\bibitem [{\citenamefont {Anisimov}\ \emph {et~al.}(1997)\citenamefont
  {Anisimov}, \citenamefont {Aryasetiawan},\ and\ \citenamefont
  {Lichtenstein}}]{AALjpcm97_767}%
  \BibitemOpen
  \bibfield  {author} {\bibinfo {author} {\bibfnamefont {V.~I.}\ \bibnamefont
  {Anisimov}}, \bibinfo {author} {\bibfnamefont {F.}~\bibnamefont
  {Aryasetiawan}}, \ and\ \bibinfo {author} {\bibfnamefont {A.~I.}\
  \bibnamefont {Lichtenstein}},\ }\href
  {http://stacks.iop.org/0953-8984/9/i=4/a=002} {\bibfield  {journal} {\bibinfo
   {journal} {J. Phys.: Condens. Matter}\ }\textbf {\bibinfo {volume} {9}},\
  \bibinfo {pages} {767} (\bibinfo {year} {1997})}\BibitemShut {NoStop}%
\bibitem [{\citenamefont {Ylvisaker}\ \emph {et~al.}(2009)\citenamefont
  {Ylvisaker}, \citenamefont {Pickett},\ and\ \citenamefont
  {Koepernik}}]{PhysRevB.79.035103}%
  \BibitemOpen
  \bibfield  {author} {\bibinfo {author} {\bibfnamefont {E.~R.}\ \bibnamefont
  {Ylvisaker}}, \bibinfo {author} {\bibfnamefont {W.~E.}\ \bibnamefont
  {Pickett}}, \ and\ \bibinfo {author} {\bibfnamefont {K.}~\bibnamefont
  {Koepernik}},\ }\href {\doibase 10.1103/PhysRevB.79.035103} {\bibfield
  {journal} {\bibinfo  {journal} {Phys. Rev. B}\ }\textbf {\bibinfo {volume}
  {79}},\ \bibinfo {pages} {035103} (\bibinfo {year} {2009})}\BibitemShut
  {NoStop}%
\bibitem [{\citenamefont {Dudarev}\ \emph {et~al.}(1998)\citenamefont
  {Dudarev}, \citenamefont {Botton}, \citenamefont {Savrasov}, \citenamefont
  {Humphreys},\ and\ \citenamefont {Sutton}}]{DBSHSprb98_1505}%
  \BibitemOpen
  \bibfield  {author} {\bibinfo {author} {\bibfnamefont {S.~L.}\ \bibnamefont
  {Dudarev}}, \bibinfo {author} {\bibfnamefont {G.~A.}\ \bibnamefont {Botton}},
  \bibinfo {author} {\bibfnamefont {S.~Y.}\ \bibnamefont {Savrasov}}, \bibinfo
  {author} {\bibfnamefont {C.~J.}\ \bibnamefont {Humphreys}}, \ and\ \bibinfo
  {author} {\bibfnamefont {A.~P.}\ \bibnamefont {Sutton}},\ }\href {\doibase
  10.1103/PhysRevB.57.1505} {\bibfield  {journal} {\bibinfo  {journal} {Phys.
  Rev. B}\ }\textbf {\bibinfo {volume} {57}},\ \bibinfo {pages} {1505}
  (\bibinfo {year} {1998})}\BibitemShut {NoStop}%
\bibitem [{Note2()}]{Note2}%
  \BibitemOpen
  \bibinfo {note} {\protect \autoref {fig:pol} is generated from 345 structures
  and performing self-consistent total energy calculations; linear
  interpolation of the computed orbital polarizations yields the full
  rotation--BD orbital polarization surfaces. We apply a two-dimensional
  moving-average filter to smooth the data for clarity.}\BibitemShut {Stop}%
\bibitem [{Note3()}]{Note3}%
  \BibitemOpen
  \bibinfo {note} {When only the breathing distortion is present, the compound
  becomes insulating at high BD amplitudes ($\geq 0.13$ \r A{}). If the BD is
  absent, rotations alone do not induce a metal-insulator transition \cite
  {CRprb12_195144}. At fixed BD, however, there are some dependencies of the
  electronic band gap on the rotation sense.}\BibitemShut {Stop}%
\bibitem [{Note4()}]{Note4}%
  \BibitemOpen
  \bibinfo {note} {Identical conclusions are obtained from considering $\theta
  _{xy}$ instead of $\theta _{z}$, \protect \emph {i.e.}, when the breathing
  distortion is coupled with $a^-a^-c^0$ rotations, $\theta _{xy}$ is an
  increasing monotonic function of $\theta _{z}$.}\BibitemShut {Stop}%
\bibitem [{\citenamefont {Rondinelli}\ \emph {et~al.}(2012)\citenamefont
  {Rondinelli}, \citenamefont {May},\ and\ \citenamefont
  {Freeland}}]{Rondinelli/May/Freeland:2012}%
  \BibitemOpen
  \bibfield  {author} {\bibinfo {author} {\bibfnamefont {J.~M.}\ \bibnamefont
  {Rondinelli}}, \bibinfo {author} {\bibfnamefont {S.~J.}\ \bibnamefont {May}},
  \ and\ \bibinfo {author} {\bibfnamefont {J.~W.}\ \bibnamefont {Freeland}},\
  }\href {\doibase 10.1557/mrs.2012.49} {\bibfield  {journal} {\bibinfo
  {journal} {MRS Bulletin}\ }\textbf {\bibinfo {volume} {37}},\ \bibinfo
  {pages} {261} (\bibinfo {year} {2012})}\BibitemShut {NoStop}%
\bibitem [{\citenamefont {Berger}\ \emph {et~al.}(2011)\citenamefont {Berger},
  \citenamefont {Fennie},\ and\ \citenamefont {Neaton}}]{BFNprl11_146804}%
  \BibitemOpen
  \bibfield  {author} {\bibinfo {author} {\bibfnamefont {R.~F.}\ \bibnamefont
  {Berger}}, \bibinfo {author} {\bibfnamefont {C.~J.}\ \bibnamefont {Fennie}},
  \ and\ \bibinfo {author} {\bibfnamefont {J.~B.}\ \bibnamefont {Neaton}},\
  }\href {\doibase 10.1103/PhysRevLett.107.146804} {\bibfield  {journal}
  {\bibinfo  {journal} {Phys. Rev. Lett.}\ }\textbf {\bibinfo {volume} {107}},\
  \bibinfo {pages} {146804} (\bibinfo {year} {2011})}\BibitemShut {NoStop}%
\bibitem [{\citenamefont {Rondinelli}\ and\ \citenamefont
  {Coh}(2011)}]{Rondinelli/Coh:2011}%
  \BibitemOpen
  \bibfield  {author} {\bibinfo {author} {\bibfnamefont {J.~M.}\ \bibnamefont
  {Rondinelli}}\ and\ \bibinfo {author} {\bibfnamefont {S.}~\bibnamefont
  {Coh}},\ }\href {\doibase 10.1103/PhysRevLett.106.235502} {\bibfield
  {journal} {\bibinfo  {journal} {Phys. Rev. Lett.}\ }\textbf {\bibinfo
  {volume} {106}},\ \bibinfo {pages} {235502} (\bibinfo {year}
  {2011})}\BibitemShut {NoStop}%
\bibitem [{\citenamefont {Dup\'e}\ \emph {et~al.}(2011)\citenamefont {Dup\'e},
  \citenamefont {Prosandeev}, \citenamefont {Geneste}, \citenamefont {Dkhil},\
  and\ \citenamefont {Bellaiche}}]{PhysRevLett.106.237601}%
  \BibitemOpen
  \bibfield  {author} {\bibinfo {author} {\bibfnamefont {B.}~\bibnamefont
  {Dup\'e}}, \bibinfo {author} {\bibfnamefont {S.}~\bibnamefont {Prosandeev}},
  \bibinfo {author} {\bibfnamefont {G.}~\bibnamefont {Geneste}}, \bibinfo
  {author} {\bibfnamefont {B.}~\bibnamefont {Dkhil}}, \ and\ \bibinfo {author}
  {\bibfnamefont {L.}~\bibnamefont {Bellaiche}},\ }\href {\doibase
  10.1103/PhysRevLett.106.237601} {\bibfield  {journal} {\bibinfo  {journal}
  {Phys. Rev. Lett.}\ }\textbf {\bibinfo {volume} {106}},\ \bibinfo {pages}
  {237601} (\bibinfo {year} {2011})}\BibitemShut {NoStop}%
\bibitem [{\citenamefont {Lv}\ \emph {et~al.}(2011)\citenamefont {Lv},
  \citenamefont {Li}, \citenamefont {Wang}, \citenamefont {Han}, \citenamefont
  {Liu}, \citenamefont {Liu}, ,\ and\ \citenamefont
  {Meng}}]{LLWHLLMapl11_202110}%
  \BibitemOpen
  \bibfield  {author} {\bibinfo {author} {\bibfnamefont {S.}~\bibnamefont
  {Lv}}, \bibinfo {author} {\bibfnamefont {H.}~\bibnamefont {Li}}, \bibinfo
  {author} {\bibfnamefont {Z.}~\bibnamefont {Wang}}, \bibinfo {author}
  {\bibfnamefont {L.}~\bibnamefont {Han}}, \bibinfo {author} {\bibfnamefont
  {Y.}~\bibnamefont {Liu}}, \bibinfo {author} {\bibfnamefont {X.}~\bibnamefont
  {Liu}}, , \ and\ \bibinfo {author} {\bibfnamefont {J.}~\bibnamefont {Meng}},\
  }\href {\doibase 10.1063/1.3662859} {\bibfield  {journal} {\bibinfo
  {journal} {Appl.Phys. Lett.}\ }\textbf {\bibinfo {volume} {99}},\ \bibinfo
  {pages} {202110} (\bibinfo {year} {2011})}\BibitemShut {NoStop}%
\bibitem [{\citenamefont {Rondinelli}\ and\ \citenamefont
  {Spaldin}(2011)}]{RSav11_3363}%
  \BibitemOpen
  \bibfield  {author} {\bibinfo {author} {\bibfnamefont {J.~M.}\ \bibnamefont
  {Rondinelli}}\ and\ \bibinfo {author} {\bibfnamefont {N.~A.}\ \bibnamefont
  {Spaldin}},\ }\href {\doibase 10.1002/adma.201101152} {\bibfield  {journal}
  {\bibinfo  {journal} {Advanced Materials}\ }\textbf {\bibinfo {volume}
  {23}},\ \bibinfo {pages} {3363} (\bibinfo {year} {2011})}\BibitemShut
  {NoStop}%
\bibitem [{\citenamefont {Benckiser}\ \emph {et~al.}(2011)\citenamefont
  {Benckiser}, \citenamefont {Haverkort}, \citenamefont {BrÃ¼ck}, \citenamefont
  {Goering}, \citenamefont {Macke}, \citenamefont {FraÃ±Ã³}, \citenamefont
  {Yang}, \citenamefont {Andersen}, \citenamefont {Cristiani}, \citenamefont
  {Habermeier}, \citenamefont {Boris}, \citenamefont {Zegkinoglou},
  \citenamefont {Wochner}, \citenamefont {Kim}, \citenamefont {Hinkov},\ and\
  \citenamefont {Keimer}}]{BHBGFYACHBZWKHKnm11_189}%
  \BibitemOpen
  \bibfield  {author} {\bibinfo {author} {\bibfnamefont {A.}~\bibnamefont
  {Benckiser}}, \bibinfo {author} {\bibfnamefont {M.}~\bibnamefont
  {Haverkort}}, \bibinfo {author} {\bibfnamefont {S.}~\bibnamefont {BrÃ¼ck}},
  \bibinfo {author} {\bibfnamefont {E.}~\bibnamefont {Goering}}, \bibinfo
  {author} {\bibfnamefont {S.}~\bibnamefont {Macke}}, \bibinfo {author}
  {\bibfnamefont {A.}~\bibnamefont {FraÃ±Ã³}}, \bibinfo {author} {\bibfnamefont
  {X.}~\bibnamefont {Yang}}, \bibinfo {author} {\bibfnamefont {O.}~\bibnamefont
  {Andersen}}, \bibinfo {author} {\bibfnamefont {G.}~\bibnamefont {Cristiani}},
  \bibinfo {author} {\bibfnamefont {H.-U.}\ \bibnamefont {Habermeier}},
  \bibinfo {author} {\bibfnamefont {A.~V.}\ \bibnamefont {Boris}}, \bibinfo
  {author} {\bibfnamefont {I.}~\bibnamefont {Zegkinoglou}}, \bibinfo {author}
  {\bibfnamefont {P.}~\bibnamefont {Wochner}}, \bibinfo {author} {\bibfnamefont
  {H.-J.}\ \bibnamefont {Kim}}, \bibinfo {author} {\bibfnamefont
  {V.}~\bibnamefont {Hinkov}}, \ and\ \bibinfo {author} {\bibfnamefont
  {B.}~\bibnamefont {Keimer}},\ }\href@noop {} {\bibfield  {journal} {\bibinfo
  {journal} {Nat. Mat.}\ }\textbf {\bibinfo {volume} {10}},\ \bibinfo {pages}
  {189} (\bibinfo {year} {2011})}\BibitemShut {NoStop}%
\bibitem [{\citenamefont {Momma}\ and\ \citenamefont
  {Izumi}(2008)}]{MIjac08_653}%
  \BibitemOpen
  \bibfield  {author} {\bibinfo {author} {\bibfnamefont {K.}~\bibnamefont
  {Momma}}\ and\ \bibinfo {author} {\bibfnamefont {F.}~\bibnamefont {Izumi}},\
  }\href {http://scripts.iucr.org/cgi-bin/paper?S0021889808012016} {\bibfield
  {journal} {\bibinfo  {journal} {J. Appl. Cryst.}\ }\textbf {\bibinfo {volume}
  {41}},\ \bibinfo {pages} {653} (\bibinfo {year} {2008})}\BibitemShut
  {NoStop}%
\end{thebibliography}
%

\end{document}